\documentclass[preprint,aps,superscriptaddress,nofootinbib,showpacs]{revtex4-1}
\usepackage{graphicx}
\usepackage{epsfig}
\usepackage{bm}
\usepackage{amssymb}
\usepackage{wrapfig}
\usepackage[usenames, dvipsnames]{color}

\newcommand{\asla}{\ooalign{\hfil/\hfil\crcr{$a$}}}

\newcommand{\Qsla}{\ooalign{\hfil/\hfil\crcr{Q}}}

\newcommand{\el}{{\cal L}}
\newcommand{\cM}{{\cal M}}
\newcommand{\cO}{{\cal O}}

\newcommand{\psibar}{\mbox{$\overline{\psi}$}}

\newcommand{\tldF}{\mbox{${\tilde F}$}}

\newcommand{\psitld}{\mbox{${\tilde{\psi}}$}}

\newcommand{\vq}{\mbox{$\bm{q}$}}
\newcommand{\vbr}{\mbox{$\bm{r}$}}

\newcommand{\vB}{\mbox{$\bm{B}$}}

\newcommand{\vlambda}{\mbox{$\bm{\lambda}$}}
\newcommand{\vsigma}{\mbox{$\bm{\sigma}$}}

\begin{document}

\title{Quantum Field Theoretic Treatment of 
Pion Production via Proton Synchrotron Radiation in 
Strong Magnetic Fields: Effects of  Landau Levels}

\author{Tomoyuki~Maruyama}
\affiliation{College of Bioresource Sciences,
Nihon University,
Fujisawa 252-8510, Japan}
\affiliation{Advanced Science Research Center,
Japan Atomic Energy Research Institute, Tokai 319-1195, Japan}
\affiliation{National Astronomical Observatory of Japan, 2-21-1 Osawa, 
Mitaka, Tokyo 181-8588, Japan}

\author{Myung-Ki Cheoun}
\affiliation{Department of Physics, Soongsil University, Seoul,
156-743, Korea}
\affiliation{National Astronomical Observatory of Japan, 2-21-1 Osawa, 
Mitaka, Tokyo 181-8588, Japan}

\author{Toshitaka~Kajino}
\affiliation{National Astronomical Observatory of Japan, 2-21-1 Osawa, 
Mitaka, Tokyo 181-8588, Japan}
\affiliation{Department of Astronomy, Graduate School of Science, 
University of Tokyo, Hongo 7-3-1, Bunkyo-ku, Tokyo 113-0033, Japan}

\author{Yongshin Kwon}
\affiliation{Department of Physics, Soongsil University, Seoul,
156-743, Korea}

\author{Grant J. Mathews}
\affiliation{Center of Astrophysics, Department of Physics,
University of Notre Dame, Notre Dame, IN 46556, USA}

\author{Chung-Yeol Ryu}
\affiliation{Department of Physics, Soongsil University, Seoul, 156-743, Korea}

\date{\today}

\pacs{95.85.Ry,24.10.Jv,97.60.Jd,}

\begin{abstract}
We study pion production from proton synchrotron radiation in the
 presence of strong  magnetic fields.
We derive the exact proton propagator from the Dirac equation in a strong
 magnetic field by explicitly including the anomalous magnetic moment.
In this exact quantum-field approach the magnitude of pion synchrotron emission
turns out to be much smaller than that obtained in the 
 semi-classical approach.
However, we also find that the anomalous magnetic moment of the proton 
greatly enhances the production rate about by two order magnitude.
\end{abstract}

\maketitle


\newpage

\section{Introduction}

Magnetic fields in neutron stars play an important role
in the interpretation of many observed phenomena.
Indeed, strongly magnetized neutron stars (dubbed  {\it magnetars}
\cite{pac92,mag3}) hold the key to understanding the asymmetry in
supernova (SN) remnants and the  still
unresolved mechanism for non-spherical SN explosions.
Such strong magnetic fields are also closely related to the  unknown
origin of the kick velocity~\cite{rothchild94} that proto-neutron stars
(PNSs) receive at  birth.

It is widely accepted that soft gamma repeaters (SGRs)
and anomalous X-ray pulsars (AXPs) correspond to magnetars
\cite{Mereghetti08}, and that the associate strong magnetic fields have 
a significant role in production high energy photons.
Furthermore, short duration gamma-ray bursts (GRBs) may arise from highly
magnetized neutron stars \cite{Soderberg06} or mergers
of binary neutron stars \cite{Gehrels05, Bloom026, Tanvit13},
and the most popular theoretical
models for the long-duration GRBs \cite{MacFadyen99, MacFadyen01, Harikae09, Harikae10}
invoke magnetized accretion disks around neutron stars or rotating
black holes (collapsars) for their central engines.
Such magnetars (or black holes with strong magnetic fields)
have also been proposed \cite{Hillas84,Aarons03} as
an acceleration site for ultra high-energy (UHE) cosmic rays
(UHECRs) and a possible association between 
magnetar flares \cite{Ioka05} and UHECRs has also been observed.

When a particle is accelerated in an external field 
or in collision with another particle, it can emit quanta corresponding 
to the field with which the particle interacts.
Synchrotron radiation can be produced by high-energy
protons accelerated in an environment containing a strong magnetic 
field.
This process has been proposed as a source for high-energy photons
in the GeV $-$ TeV range
\cite{Gupta07,Boettcher98,Totani98,Fragile04,Asano07,Asano09},
possibly in association with GRBs.
Synchrotron emission, however, can occur through any
quanta that may couple to an accelerated particle.
Since protons strongly couple to meson fields, a high-energy proton can 
also radiate pions and other mesons, as well as photons.

In fact, the meson-nucleon couplings are about 100 times larger than the
photon-nucleon coupling, and the meson production process is expected to 
exceed photon synchrotron emission in the high energy regime. 
For example, Refs. \cite{Ginzburg65a,Ginzburg65b,Zharkov65,TK99,BDK95}
addressed the possibility of $\pi^0$ emission from  a proton in a strong
magnetic field. However, these calculations were  performed
approximately; i.e. in a semi-classical way and an approximate
quantum-mechanical treatment of the proton transitions among
the Landau levels of the strong magnetic field.

\bigskip

In the semi-classical approach for the production of synchrotron radiation, 
the magnetic field strength is characterized by the curvature parameter, 
$\chi = E_i^2 / (m^3 R_c)$, given in terms of  the incident particle mass  
$m$, its energy $E_i$, and the curvature radius $R_c$. 
For $\chi << 1 ~( R_c >>1)$, the
radiation can be treated in a classical way. 
For example, in a gravitation field or a relatively small electro-magnetic 
field, charged particles may have such large curvature radii. 
However, for $\chi >> 1 ~( R_c << 1)$ {\it i.e.}, 
very high energetic particles propagating 
in the very small curvature radius of a strong magnetic field, 
quantum effects must be taken into account. 
Since we consider a proton propagating in a strong magnetic field, which may
imply a small Lamor radius $R_L = E / (e B)$, the
radiation should be also treated in a quantum mechanical way along with 
the Landau quantization by the magnetic field. 

For proton propagation in a strong magnetic field, the value of $\chi$ 
can be given as
\begin{equation}
\chi = {E_i^2 \over {M_p^3 R_L}} = {{e E_i B} \over {M_p^3}} =
 {E_i \over M_p} ( {B_{\perp} \over B_{cr}} )~,
\end{equation}
where $B_{cr} = M_p^2 / e$ and $B_{\perp}$ is the magnetic field
perpendicular to the proton momentum direction.  
For illustration, we assume a 1 $\sim$ 5 GeV proton propagating in a strong 
magnetic field with $B \sim 10^{18}$G, for which $\chi \sim 0.01$. 
Through analysis of the pion emission based on
a quantum field theory treatment, we show that the pion emission can be
comparable to the photon emission. 
In particular, we demonstrate that the anomalous magnetic
moment of the charged particle plays a vital roles in the pion
synchrotron emission.  

\bigskip

In this work, we exploit the Green's function method for the propagation 
of protons in a strong magnetic field.
This approach has not previously been applied to derive pion 
synchrotron emission. 
For the pion-nucleon coupling, the $p$-wave interaction is dominant, and
the pion emission amplitude is mainly proportional to 
$<\psibar_N \vsigma \cdot \vq \psi_N> $, with ${\bm q}$ being the emitted pion
momentum \footnote{ Note: this fact is independent of the choice of whether 
one has PS- or PV-coupling.}. 
Because the pion  emission along the direction of the magnetic field 
is not allowed by the conservation of energy-momentum, 
$<\psibar_N \sigma_{\pm} \psi_N>$ has a dominant contribution.
Hence, the spin-flip contributions necessary for pseudo-scalar emission 
are significant. 

Section 2 is devoted to an introduction of our theoretical formalism  
based upon the Green's function method. 
Numerical results are presented 
in section 3 along with detailed discussions. 
A summary and conclusions are presented in section 4.   

\newpage

\section{Formalism }

\subsection{Proton Green's Function}

We assume a uniform magnetic field along the $z$-direction as
$\vB = (0,0,B)$ and take the electro-magnetic vector potential $A^{\mu}$ to be
$A = (0, 0, x B, 0)$ at the position $\vbr \equiv (x, y, z)$ .

The relativistic proton wave function $\psitld$ is obtained
from the following Dirac equation:
\begin{equation}
\left[ \gamma_\mu \cdot (i \partial^\mu - e A^\mu) - m_N
- \frac{e \kappa_p}{2 m_N} \sigma_{\mu \nu}
(\partial^\mu A_\nu - \partial^\nu A^\mu ) \right]
\psitld (x) = 0 ,
\label{DirEq}
\end{equation}
where $m_N$ is the proton mass, $\kappa_p$ is the proton anomalous magnetic
moment (AMM), and $e$ is the elementary charge.

Here, we scale all variables with $\sqrt{eB}$ as
$X_\mu  = \sqrt{eB} x_\mu$ and $M_N = m_N /\sqrt{eB}$, and we write the
wave function, e.g.
\begin{equation}
 \psi(X) = \left( \begin{array}{c}
\lambda_1 f_{n+1} (X- P_y)  \\ \lambda_2 ~ f_{n} (X- P_y)~  \\
\lambda_3 f_{n+1} (X- P_y)  \\ \lambda_4 ~ f_{n} (X- P_y)~
		  \end{array}\right) e^{i (P_y X_2 +P_z X_3 - i E X_0) } .
\end{equation}
Eq.~(\ref{DirEq}) then leads the following characteristic equation:
\begin{eqnarray}
&& \left[ E \gamma_0 +  \sqrt{2(n + 1)} \gamma^2 - P_z \gamma^3
- (\kappa_p/M_N) \Sigma_z - M_N \right] \vlambda
= 0
\label{ChrEq}
\end{eqnarray}
with $\vlambda = (\lambda_1, \lambda_2, \lambda_3, \lambda_4)$ and
$\Sigma_z = diag(1,-1,1,-1)$.

Here, the Landau level $n$ starts from $n=-1$ when $s=1$
and  from $n=0$ when $s=-1$.
We then redefine the Landau number
$n_L$ as $n_L = n + (1 + s)/2$.
By solving Eq.~(\ref{ChrEq}), we then obtain the energy eigenvalues as
\begin{equation}
E(n_L, P_z, s) = \pm \sqrt{ P_z^2 + (\sqrt{2n_L + 1 - s + M_N^2}
- s \kappa_p/M_N)^2}.
\end{equation}

The proton Green's Function $G$ in a magnetic field is written as
\begin{eqnarray}
G &=& \sum_{n_L=0} \sum_{s= \pm 1} \tldF(X) \left[
\frac{\rho_M^{(+)}(n_L,s,P_z)}{P_0 - E(n_L,s,P_z) + i \delta}
+ \frac{\rho_M^{(-)}(n_L.x.P_z)}{P_0 + E(n_L,s,P_z) + i \delta}
\right]  \tldF(X^\prime)  
\end{eqnarray}
with
\begin{eqnarray}
\tldF
&=& f_{n_L+\frac{1-s}{2}} (X-P_y)  \frac{1 + \Sigma_z }{2}
+  f_{n_L-\frac{1+s}{2}}(X-P_y)  \frac{1 - \Sigma_z }{2}  ,
\\
\rho_M^{(+)} &=&
\frac{1}{4 E }
\left[ E \gamma_0 + \sqrt{2n_L + 1 - s} \gamma^2 - P_z \gamma^3
 + M_N + (\kappa_p/M_N) \Sigma_z \right]
\nonumber \\ && \quad \times
\left( 1 + \frac{s \kappa_p/M_N }{\sqrt{ 2n_L+1-s +M_N^2} }
+ \gamma_5 \asla \right) ,
\\
\rho_M^{(-)} &=&
\frac{1}{4 E }
\left[ E \gamma_0 - \sqrt{2n_L + 1 - s} \gamma^2 + P_z \gamma^3
- M_N - (\kappa_p/M_N) \Sigma_z  \right]
\nonumber \\ && \quad \times
\left( 1 + \frac{s \kappa_p/M_N}{\sqrt{ 2n_L+1-s +M_N^2} }
- \gamma_5 \asla \right) ,
\end{eqnarray}
where $E_F$ is the Fermi energy, and the spin vector $a$ is defined as
\begin{equation}
 a (p,s) = \left( \frac{sP_z}{ \sqrt{ 2n_L + 1 - s + M_N^2}} , 0, 0,
 \frac{sE}{ \sqrt{ 2n_L + 1 - s + M_N^2}} \right).
\end{equation}

\subsection{Pion Production }

In this subsection we consider the pion production rate in the presence
of a strong magnetic field.
We start the calculation from the following pseudo-vector coupling
interaction Lagrangian density as
\begin{equation}
 \el = \frac{ if_{\pi} }{m_{\pi}} \psi \gamma_5 \gamma_\mu \tau_a \psi
\partial^{\mu} \phi_{a} ,
\end{equation}
where $f_\pi$ is the pseudo-vector pion-nucleon coupling constant,
$m_\pi$ is the pion mass, and $\phi$ is the pion field. 
We then calculate the pion decay rate from an initial proton with
$n_L=n_i$, $s=s_i$ and $P_z = P_{iz}$ to a final proton 
with $n_L=n_f$, $s=s_f$ and $P_z = P_{fz}$.
The pion momentum scaled by $\sqrt{eB}$ is written as
$Q=(E_{\pi}, 0, Q_T, Q_z)$, where  without loss of generality 
the transverse pion momentum is assumed to be directed along the $y$-axis.

Using the proton propagator, the proton self-energy from 
one-pion exchange can be written as
\begin{equation}
\Sigma_\pi (R_1, R_2) =  i \left( \frac{f_\pi}{M_\pi} \right)^2
\gamma_{\mu} \gamma_5 \tau_a  G (R_1, R_2) \tau_a \gamma_{\nu} \gamma_5
\partial^\mu \partial^\nu \Delta_{\pi}(R_1-R_2) ,
\nonumber
\end{equation}
and the decay width is calculated from the imaginary part of the expectation
value of the self-energy  as
\begin{equation}
\Gamma_\pi = - {\rm Im}
\int d^3 R_1 d^4 R_2 \psibar_i (R_1) \Sigma_{\pi} (R_1, R_2) \psi_i (R_2) ,
\end{equation}
where $\Delta_{\pi}$ is the pion propagator.

By performing a Fourier transformation, we can obtain  
the differential decay width of the proton as
\begin{equation}
\frac{d^3 \Gamma_{p \pi} / \sqrt{eB}}{d Q^3} =
\frac{1}{8 \pi^2 E_\pi}  \left( \frac{f_\pi}{M_{\pi}} \right)^2
\sum_{n_f,s_f}  \frac{\delta(E_f + E_{\pi} - E_{i})}{4 E_i E_f} W_{if} ,
\label{dfWid}
\end{equation}
with
\begin{equation}
W_{if} = 4 E_i E_f {\rm Tr} \left\{ \rho_M^{(+)} (n_i, s_i, P_z) \cO_{\pi}
\rho_M^{(+)} (n_f, s_f, P_z - Q_z) \cO_{\pi}^{\dagger}
 \right\} ,
\end{equation}
where $M_\pi = m_{\pi} /\sqrt{eB}$,  and
\begin{eqnarray}
\cO_{\pi}
& =& \int d X \tldF(n_i,s_i,X+Q_T/2) \gamma_5 \Qsla \tldF(n_f,s_f,X-Q_T/2)
\nonumber \\
&=&
\gamma_5 \left\{ \left[
\cM \left( n_i + \frac{1-s_i}{2}, n_f+\frac{1-s_f}{2} \right)
\frac{1 + \Sigma_z}{2}
\right.\right. \nonumber \\ && \left. \quad \quad \quad
+  \cM \left( n_i - \frac{1+s_i}{2}, n_f - \frac{1+s_f}{2} \right)
\frac{1 - \Sigma_z}{2} \right]
\left[ \gamma_0 Q_0 - \gamma^3 Q_z \right]
\nonumber \\ &&  \quad
- \left[ \cM \left( n_i + \frac{1-s_1}{2}, n_f-\frac{1+s_f}{2} \right)
 \frac{1 + \Sigma_z}{2}
\right. \nonumber \\ && \left.\left. \quad \quad \quad
+  \cM \left( n_i - \frac{1+s_i}{2}, n_f+ \frac{1-s_f}{2} \right)
\frac{1 - \Sigma_z}{2} \right] \gamma^2 Q_y  \right\} .
\end{eqnarray}
In the above equation, $\cM(n_1, n_2)$ is defined as
\begin{eqnarray}
 \cM (n_1,n_2)  & =&
\int d x f_{n_1}\left( x+\frac{Q_y}{2} \right)
f_{n_2} \left( x-\frac{Q_y}{2} \right) 
\nonumber \\ &=&
(2^{n_1+n_2} \pi n_1 ! n_2 !)^{-1/2} e^{-Q_T^2/4 }\int  dx e^{-x^2}
H_{n_1} \left( x + \frac{Q_T}{2} \right)
H_{n_2} \left( x - \frac{Q_T}{2} \right) 
\nonumber \\
&=&
\sqrt{ \frac{n_1!}{n_2! } }
\left( - \frac{Q_T}{\sqrt{2} } \right)^{n_2-n_1}
e^{-\frac{Q_T^2}{4} } L^{n_2-n_1}_{n_1} \left( \frac{Q_T^2}{2} \right)
\quad\quad (n_1 \le n_2 ) ,
\nonumber \\ &=&
\sqrt{ \frac{n_2!}{n_1! } }
~\left( \frac{Q_T}{\sqrt{2} } \right)^{n_1 - n_2} ~
e^{-\frac{Q_T^2}{4} } L^{n_1-n_2}_{n_2} \left( \frac{Q_T^2}{2} \right)
\quad\quad~ (n_1 \ge n_2 ) ,
\label{TrStM}
\end{eqnarray}
where $H_n(x)$ is the $n$-th Hermite polynomial, and $L_n^m (x)$ is
the associated Laguerre polynomial.

\newpage
\section{Results}

In this section we show numerical results for pion emission from $1-2$
GeV protons.
In the numerical calculation we take the maximum Landau level to be
$\sim 3 \times 10^3$, which somewhat limits our calculations.
However, in order to test our model,
we choose a proton energy of 1 GeV and a strength of the magnetic field 
to be $5 \times 10^{18}$G.  
In this case the maximum Landau numbers are $n_{max}=45$ for spin  $s=-1$ and
  $n_{max}=50$ for  $s=+1$.
Pion production has been shown \cite{TK99} to exceed photon emission 
in the region,
$\chi = eB e_i/m_N^3 = 0.01 \sim 1$, in the semi-classical calculation.
The value of $\chi$, 0.069, adopted in this calculation belongs in this region.

\begin{wrapfigure}{r}{9.4cm}
\vspace{-1cm}
\begin{center}
{\includegraphics[scale=0.55]{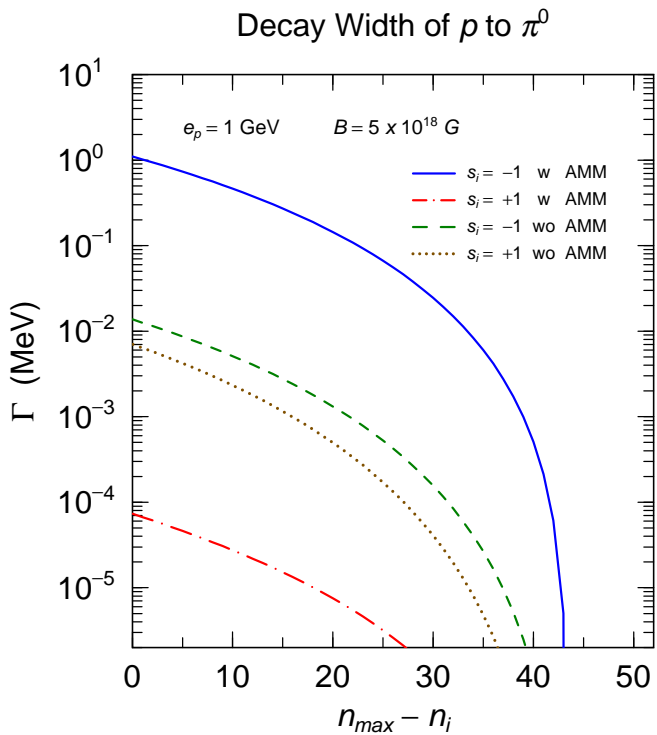}}
\caption{\small
(Color online) Decay width of a 1GeV proton for the synchrotron emission 
of $\pi$'s.
The solid and dot-dashed lines represent the decay widths of protons with
spin $s_i=-1$ and $s_i=1$, respectively.
The dashed and dotted lines indicate results with $s_i=-1$ and $s_i=1$,
respectively, for case that the anomalous magnetic moment is omitted.
}
\label{E1B5}
\end{center}
\end{wrapfigure}

In Fig.~\ref{E1B5} we show the pionic decay widths of the proton at the
Landau level $n_i$ and the spins state $s_i$ with a proton  kinetic
energy of 1~GeV. 
All results are summed over the final proton spin and Landau levels.
The solid and dot-dashed lines represent the decay widths of the proton
with spin $s_i = -1$ and $s_i=+1$, respectively.
For comparison we also plot results when the AMM is set to be zero,
$\kappa_p= 0$, with the dashed ($s_i =-1$)  and dotted lines ($s_i=+1$).

When $\kappa_p=0$, the decay width of the proton with $s_i=-1$ is
a little larger than that with $s_i=+1$, 
but these results are not significantly different.
In addition, the two states of $s = \pm 1$ are degenerate except for the
lowest Landau state, and the spin-up and spin-down states cannot be 
uniquely determined when the AMM does not exist,
Thus, this difference is not quantitatively meaningful.

On the other hand, when the AMM is included, 
the width with $s_i=-1$ is enhanced up to about a factor of 100,
while that with $s_i=+1$ is suppressed by a factor of  about 1/100.

\begin{wrapfigure}{r}{8.7cm}
\begin{center}
{\includegraphics[scale=0.45,angle=270]{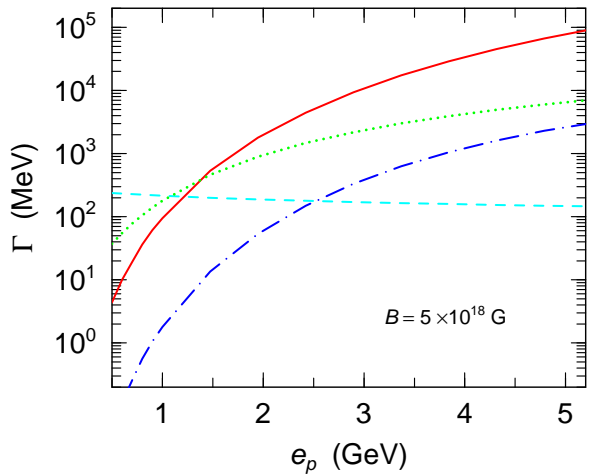}}
\caption{\small
(Color online) Pionic decay width of protons versus the proton incident energy
for $B=5 \times 10^{18}$G.
The decay width is averaged over the proton spin, and
the Landau number is taken to be a maximum.
The solid and dot-dashed lines represent the decay widths of the proton with
and without the AMM, respectively.
The dashed and dotted lines indicate the results in the semi-classical
 approaches of Refs. \cite{TK99} and in \cite{BDK95}, respectively.
}
\label{WdQCB5}
\end{center}
\end{wrapfigure}

In Fig.\ref{WdQCB5}, we compare our decay widths to those in the
semi-classical approach. 
This figure  shows the decay width averaged over the proton
spin, $\Gamma = [ \Gamma(n_{max},+1) + \Gamma (n_{max}, -1) ]/2$,
when $B=5 \times 10^{18} G$ as a function of the incident proton energy.

When the proton energy is $e_p = 1$~GeV, the semi-classical
approach gives $\Gamma_{sc} \approx 180$~ MeV in
Refs. \cite{TK99} and 210 MeV in \cite{BDK95}.
These values are close to the maximum value of the pionic width of the proton
with the AMM when $s_i=-1$.
However, the AMM is not taken into account in the semi-classical
calculations, and the results in the semi-classical approach should be
compared with our results without the AMM.
In addition, the expression of Ref.~\cite{BDK95} is derived with the
condition, $\chi \gg 1$, which is not consistent to the present condition,
$\chi \approx 0.06 - 0.5$.
So, the decay width in the microscopic approach is much smaller than that 
in the semi-classical approach for the case without the AMM.
Therefore, the AMM contribution, which is unique in the quantum
mechanical approach, is vital for pion synchrotron radiation,
particularly, in the limit of small curvature radii. 

As the proton energy increases, the decay widths in the microscopic
framework drastically increase, while those in the semi-classical
approach increase more gently.
In the semi-classical calculation one assumes that the Landau level number 
is very large. 
For example, in Ref.~\cite{BDK95}, they take the limit, 
$n_i \rightarrow \infty$. 
In Fig.\ref{WdQCB5}, on the other hand, 
the maximum Landau  is taken to be $n_{max} \approx 20 -700$, which is
too small for the semi-classical approximation.
Thus, the semi-classical calculation is not justified based upon 
the present condition.  
However, the AMM contribution turns out to significantly 
increase the pion emission in the limit of a strong magnetic field.

\begin{figure}[ht]
\begin{center}
{\includegraphics[angle=270,scale=0.61]{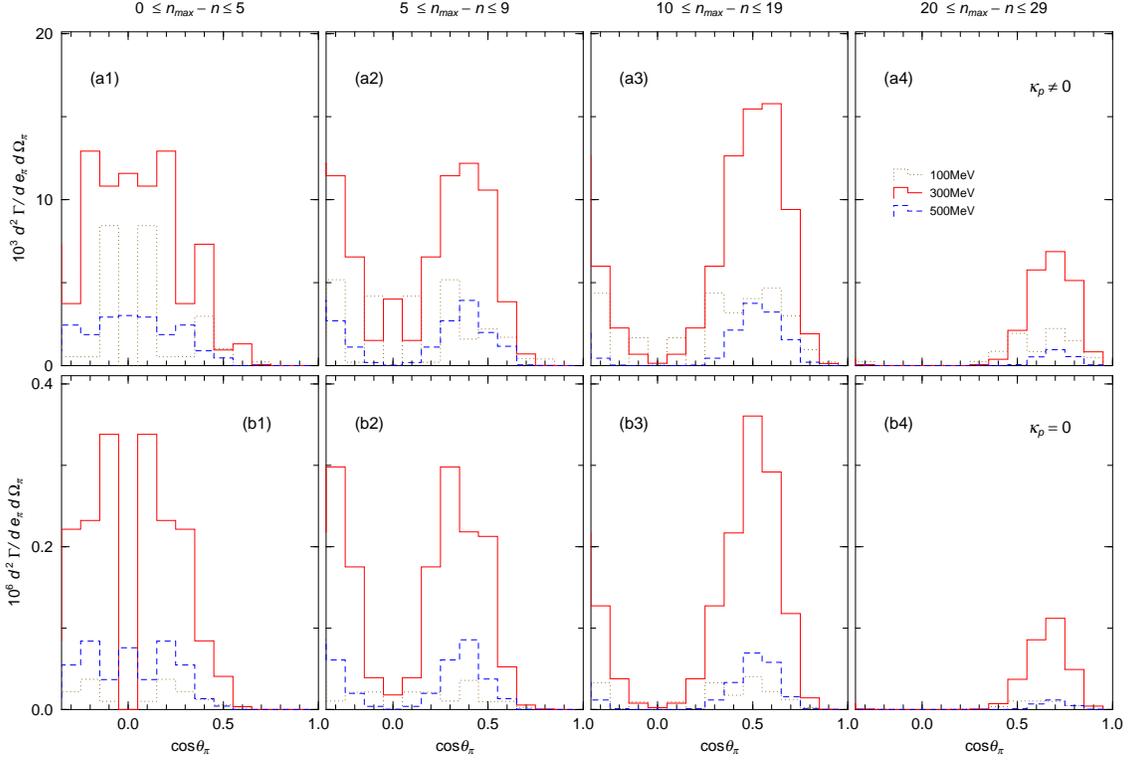}}
\caption{\small
(Color online) The differential proton pionic decay width versus the
 polar angle of pion emission.
The widths are averaged over the initial  Landau levels,
$0 \le n_{max} - n_i \le 4$  (a1,b1), $5 \le n_{max} - n_i \le 9$ (a2, b2),
$10 \le n_{max} - n_i \le 19$ (a3, b3) and
$20 \le n_{max} - n_i \le 29$ (a4, b4).
The AMM is included in the upper panels (a1$-$4),
and not included in the lower panels (b1$-$4).
The emitted pion energies are taken to be $100$~MeV (dotted lines),
300~MeV (solid lines) and 500~MeV (dashed lines).}
\label{PiAnE1B5}
\end{center}
\end{figure}

In order to understand the angular distribution of emitted pions, in
Fig.~\ref{PiAnE1B5}, 
we present the differential pionic decay widths of the proton, 
averaged over  various initial  Landau levels,
$0 \le n_{max} - n_i \le 4$  (a1,b1), $5 \le n_{max} - n_i \le 9$ (a2, b2),
$10 \le n_{max} - n_i \le 19$ (a3, b3) and
$20 \le n_{max} - n_i \le 29$ (a4, b4). 
The AMM contribution is included in the upper panels (a1$-$a4), but not included in the lower panels (b1$-$b4).
The emitted pion energies are taken to be $100$ MeV (dotted lines),
300 MeV (solid lines), 500 MeV (dashed lines).
Because the differential decay width is discrete for each angle,
we average the results over the polar angle in every angular bin with 
$\Delta \cos \theta_\pi = 0.1$.

\begin{wrapfigure}{r}{9.5cm}
\vspace{-0.5cm}
\begin{center}
{\includegraphics[scale=0.52]{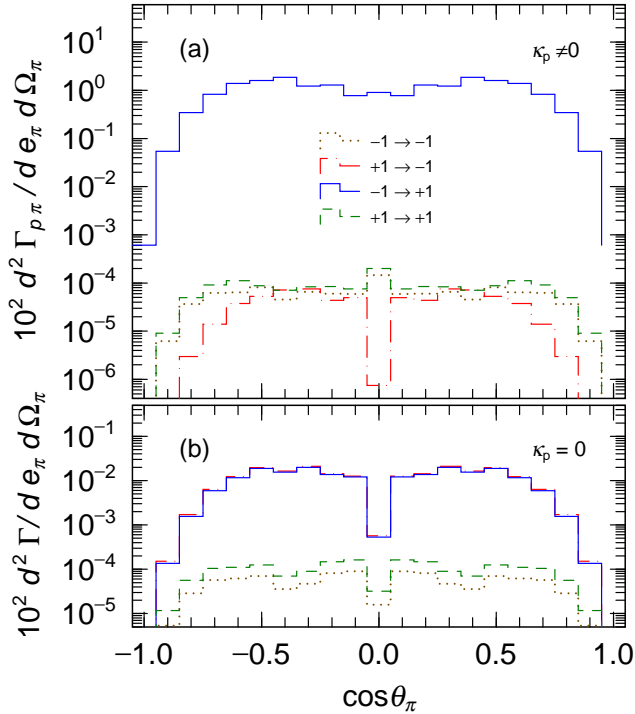}}
\caption{\small
(Color online) The differential pionic decay widths of protons with  (a)
and without (b) the AMM included. 
The widths are averaged over initial Landau numbers,
$0 < n_{max} - n_i < 9$.
The solid, dot-dashed, dashed, and dotted lines represents the results
when $s_i= - s_f= -1$,  $s_i= -s_f=1$, $s_i=s_f= 1$,
and $s_i=s_f=-1$, respectively, where $s_i(f)$ indicates the initial (final)
spin of the proton.
}
\label{WdSPE1B5}
\end{center}
\end{wrapfigure}

We should note that the angular distribution is symmetric between
$\cos \theta_\pi <0$ and $\cos \theta_\pi >0$, so that we do not plot 
results for all $\cos \theta_\pi$. 
A pion produced from a proton with a higher Landau level $n_i$
is emitted more transversely, and the AMM effect shifts the pion emission to
a more sideward direction when $0 \le n_{max} - n_i \le 4$. 
If we consider that the emitted pion decays into 2$\gamma$s, 
it could affect the relativistic beaming of the energetic gamma rays from GRBs.

Next we examine the proton-spin condition for the pionic decay width.
In Fig.~\ref{WdSPE1B5} we give the initial and final spin-dependence of
the proton differential pionic decay widths
with (a) and without (b) the AMM.
These widths are averaged over the initial Landau levels
$0 \le n_{max} - n_i \le 9$.
The solid lines represent the results when the initial spin $s_i= -1$
and the final spin $s_f=1$.
The dot-dashed, dashed and dotted line indicate cases for which 
$s_i=- s_f= 1$, $s_i= s_f=1$, and $s_i=s_f=-1$, respectively.

When $\kappa_p=0$, the contributions from the spin-flip,
$s_i = - s_f$,  are about 100 times larger than those of the
spin non-flip, $s_i = s_f$.
Semi-classically  the amplitude of the pion emission 
is proportional to $<\psibar_N \vsigma \cdot \vq \psi_N> $.
The pion  emission along the $z$-direction is not allowed  by 
energy-momentum conservation, and $<\psibar_N \sigma_{\pm} \psi_N>$
is the dominant contribution.
Here, the  spin-flip contributions become
much larger than those from the spin non-flip reaction.

When the AMM is included, only the contribution from $s_i = - s_f = 1$
is about 10,000 times larger than those of the other channels, and 
the non spin-flip contributions are not much different between those
with and without the  AMM.

As shown in Fig.~\ref{E1B5}  the AMM  increases the decay width by 100 times 
for the case of $s_i=1 = - s_f =1$, and decreases it for the case of 
$s_i =-s_f=-1$.
When $s_i=-s_f=1$, the effects of the AMM and spin-flip are synchronized, 
and they enlarge the width by up to a factor of $10^3$.
When $s_i=-s_f=1$, the two effects cancel and the width stays on
the same order as that of the non spin-flip transition.

\begin{figure}[htb]
\vspace{0.5cm}
\begin{center}
{\includegraphics[angle=270,scale=0.6]{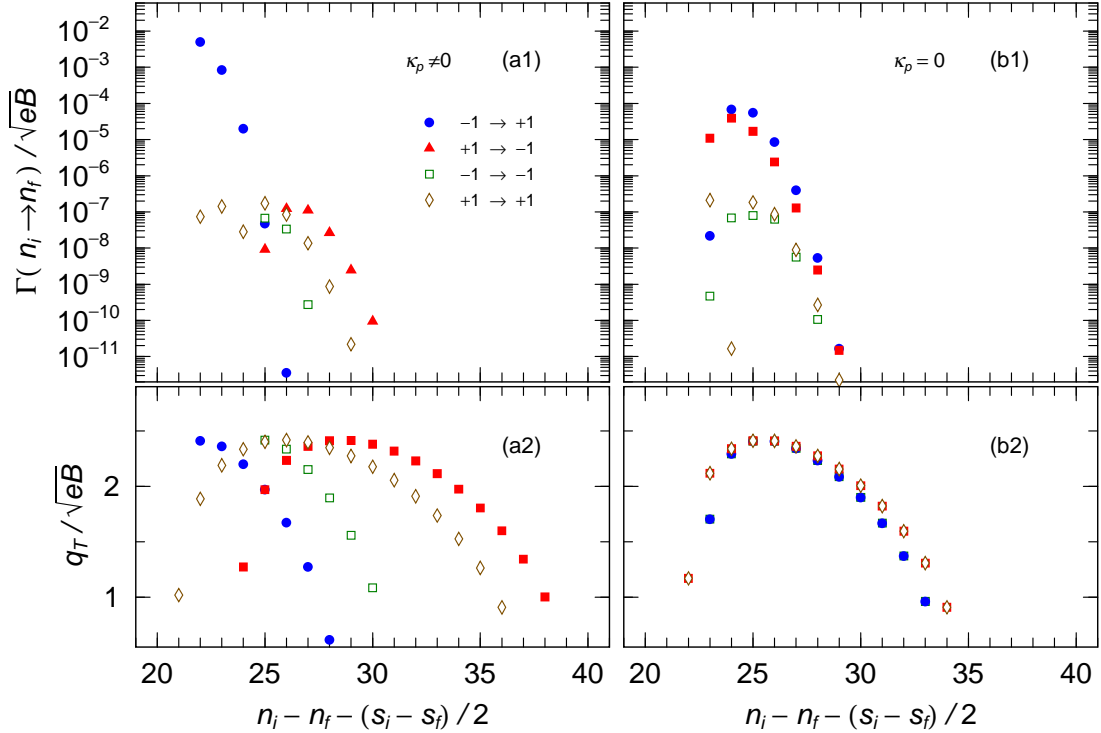}}
\caption{\small
(Color online) Pionic decay widths of protons  (a1, a2)
and the pion transverse momentum (b1, b2) as functions of
$n_i - n_f -(s_i - s_f)/2 $.
The AMM is included in the left panels (a1, b1),
and not included in the right panels (a2, b2).
The initial Landau number is fixed to be $n_i +(1 - s_i)/2 = 45$.}
\label{LndInFn}
\end{center}
\end{figure}

Next, we examine contributions from the final Landau level.
In Fig.~\ref{LndInFn} we show the width (a1 and a2) and the pion transverse 
momentum (b1, b2) as functions of $n_i - n_f - (s_i - s_f)/2$, 
where the initial Landau number is fixed to be $n_i +(1 - s_i)/2  =45$.
The results are calculated with and without the AMM in the left (a1, b1)
and right panels (a2 and b2), respectively.

Here, we note that the peak positions of the decay widths and
the pion transverse momentum are the same. 
When $\kappa_p =0$, the peak position is almost independent of
the initial and final spin.
When the AMM is included, however, the peak position of the non spin-flip
transition is the same as that without the AMM, but
 it is shifted to smaller values when
$s_i = - s_f = -1$ and to larger values when $s_i = - s_f = +1$.

\begin{figure}[bht]
\begin{center}
{\includegraphics[angle=270,scale=0.5]{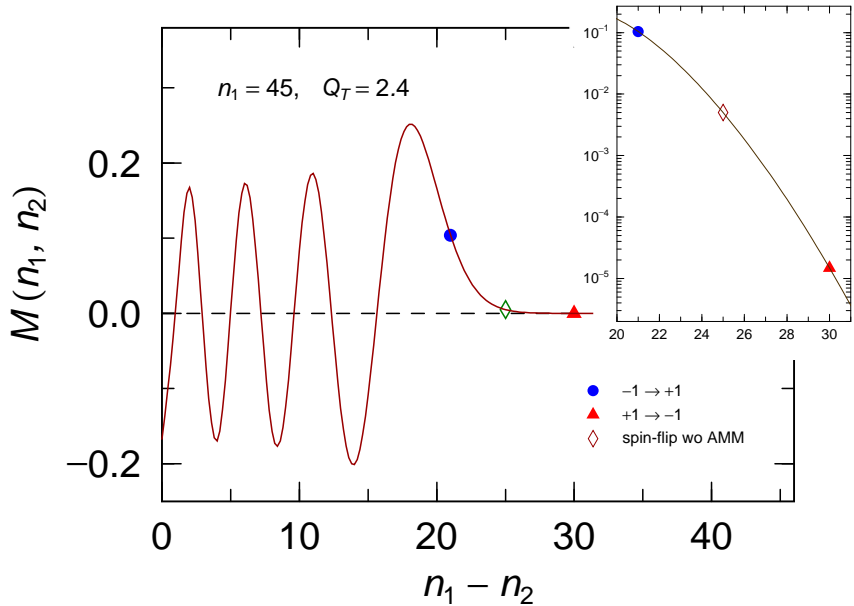}}
\caption{\small
(Color online) The transition strength function $M (n_i, n_f )$,
 Eq.~(\ref{TrStM}), when $n_i = 45 +(1 + s_i)/2$.
The solid circle and triangle show the peak positions in
the cases $s_i = - s_f =-1$ and  $s_i = - s_f =1$, and the opened diamond 
indicates the case of the spin-flip $s_i = - s_f$ when $\kappa_p =0$.
}
\label{MSt1}
\end{center}
\end{figure}

In order to study the AMM effect more clearly, we next examine
the transition strength $M(n_i, n_f)$ defined in Eq.~(\ref{TrStM}).
In Fig.~\ref{MSt1} we show the $n_f$-dependence of $M(n_i, n_f)$ when
$n_i = 45 +(1 + s_i)/2$.
The solid circle and triangle in the inset represent the peak positions in
the cases with $s_i = - s_f =-1$ and  $s_i = - s_f =1$ with the AMM included.  
The open diamond indicates the case of a spin-flip $s_i = - s_f$ 
without including AMM.

$M(n_i, n_f)$ shows an oscillating behavior, but only the strength
after the last peak contributes to the results.
In this region the strength rapidly decreases with increases of $n_i-n_f$.
A pion cannot be produced in the free kinematics, $B=0$, because of 
energy momentum conservation.
Under the influence of a magnetic field, momentum conservation is not satisfied,
so that a pion can be produced in the kinematical condition far from
the free kinematics, where $M(n_1, n_2)$ rapidly decreases.

The AMM gives a repulsive potential for $s_i=-1$ and attractive for $s_i=1$.
The transition with $s_i=s_f=-1$ introduces an additional energy to be
consumed in the pion production. 
Thus, the difference between the initial and final Landau-levels is 
shifted toward a smaller number.
In addition, the transition with $s_i=s_f=1$ reduces the production energy.
In this kinematical region a small difference between the initial and
final Landau-levels significantly changes the transition strength.
Therefore, the AMM plays the important roles of increasing greatly 
the pionic decay width when  $s_i=s_f=-1$ and to decrease it  $s_i=s_f=1$.

\begin{figure}[bht]
\begin{center}
{\includegraphics[scale=0.5]{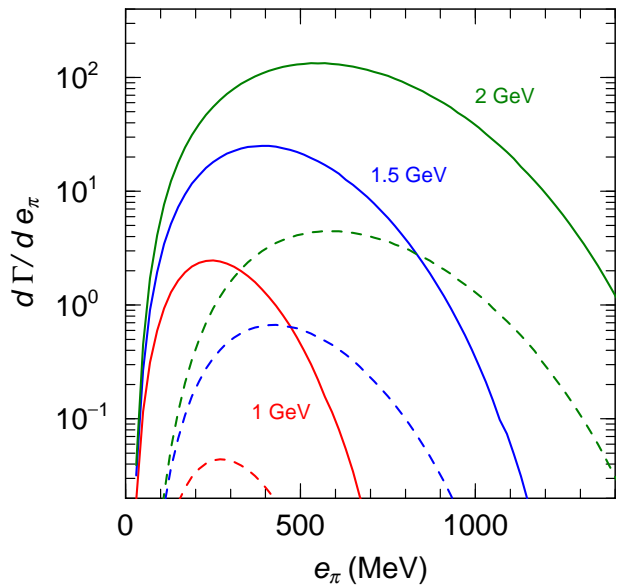}}
\caption{\small
(Color online) The dependence of the proton decay width on the pion energy 
for a magnetic field of $B=5\times10^{18}$~G.
The solid and dashed lines represent the results with and without the AMM.
The proton incident energies are taken to be 1, 1.5 and 2 GeV, respectively.}
\label{WdEn}
\end{center}
\end{figure}

Next, we study the energy spectrum of the produced pions.
We assume spherical symmetry in the momentum distribution
of the initial protons and perform the angular integration of the
differential decay width.
In Fig.~\ref{WdEn} we show the dependence of the proton decay width 
on the pion energy assuming a magnetic field $B=5 \times 10^{18}$G
and the proton energies of  $e_P = 1$, 1.5, 2GeV, respectively.
The solid and dashed lines represent the results with and without the AMM,
respectively.

As the proton energy increases, the decay width becomes larger, 
and the peak pion energy also increases.
When the AMM is included, the peak height is about 59 times larger 
than that when it is neglected for $e_p=1$GeV, and its ratios are 
36 for $e_p=1.5$GeV  and 30 for $e_p=2$GeV;
This difference between the peak height with the AMM and that without the AMM
becomes smaller with increasing proton energy.

\begin{figure}[bht]
\vspace{0.5cm}
\begin{center}
{\includegraphics[scale=0.5]{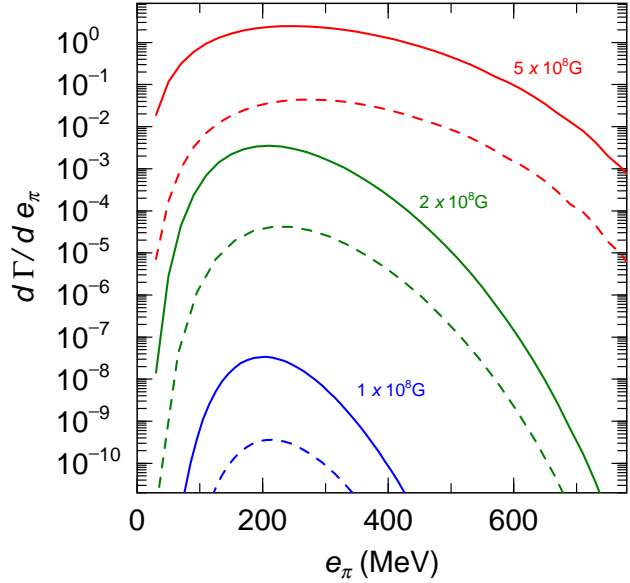}}
\caption{\small
(Color online) The pion energy dependence of the proton decay width when
the proton energy is $e_p =1$GeV.
The solid and dashed lines represent results with and without the AMM.
The magnetic fields are taken to be $1 \times 10^{18}$G,  $2 \times 10^{18}$G,
and $5 \times 10^{18}$G. 
Results increase with an increase of the magnetic field strengths.
}
\label{WdMg}
\end{center}
\end{figure}

In Fig.~\ref{WdMg} we show the dependence of the proton decay width
on the pion energy for a proton energy $e_p = 1$GeV
with magnetic fields,  $B = 2 \times 10^{18}$G,
$5 \times 10^{18}$G and $8 \times 10^{18}$G.
The solid and dashed lines represent the results with and without the AMM,
respectively.

As the magnetic field increases, the decay width becomes larger, 
and the peak pion energy also increases.
The peak height with the AMM included is about 93 times larger than 
that without the AMM for $B=1 \times 10^{18}$G,  and its ratios 
are 84 when $B=2 \times 10^{18}$G,  56 when  $B=5 \times 10^{18}$G;
the difference between the peak height with the AMM included and 
that without the AMM becomes larger as  the magnetic field  decreases.

As noted above, a pion can be produced in conditions
far from that of free kinematics.
As the magnetic field decreases, the breaking of momentum
conservation becomes larger, so that the effect of the AMM becomes more
significant.


\section{Summary}

In this work we have calculated the pion synchrotron radiation from high
energy protons propagating in strong magnetic fields in a microscopic 
quantum field theoretical framework. 
We solved the Dirac equation in a strong magnetic field and 
obtained the proton propagator from its solution. 
Then,  we derived the pionic decay width of the propagating proton 
in a fully relativistic and quantum mechanical way.

Our results turn out to be compatible to those obtained 
by classical approaches. 
In particular, we find out that the anomalous magnetic moment has a very
large effect which enlarges the emission rate by about 50 times, when the
proton energy is 1GeV and the magnetic field is $5 \times 10^{18}$G.
In actual magnetars the surface magnetic field is known to be of order 
$B \sim 10^{15}$G . 
In the present method we did not perform a calculation for such 
a magnetic field strength  because of the large number of Landau levels 
involved: a few thousand to a few million.

As the magnetic field decreases, the AMM effect becomes larger.
In a small magnetic field the decay width is very small, so that the
proton energy must be larger to produce pions;
if the semi-classical estimate is carried, the proton energy is expected
to be at least about 100 GeV, and  the maximum landau number should
be a few hundred thousand.
As the proton energy increases, on the other hand, 
the AMM effects diminish.
Though it is not easy to estimate results for $B \sim 10^{15}$G, 
one expects the AMM effect to remain.

As for future studies, since the emitted pion can decay into two gammas with
some angular dependence with respect to the magnetic field, the
secondary produced gamma ray may affect the photons from  GRBs. 
More detailed studies are necessary for further discussion of 
that additional effect. 
Since the pions as well as the photons (including vector mesons) may be
emitted in strong magnetic fields, 
the weak $Z^0$ boson can also be produced from the propagating proton. 
This boson can decay into a pair of neutrinos. 
Detailed calculations regarding the neutrino pair production 
is also in progress.

\acknowledgements
Work at NAOJ was supported in part by Grants-in-Aid 
for Scientific Research of JSPS (26105517, 24340060).
Work at the University of Notre Dame is supported
by the U.S. Department of Energy under 
Nuclear Theory Grant DE-FG02-95-ER40934.


\appendix

\section{Proton Pionic Decay Width}

In this section we show the detailed expressions of the proton pionic
decay width, Eq.~(\ref{dfWid}).
\begin{equation}
\frac{d^3 \Gamma_{p \pi} / \sqrt{eB}}{d Q^3} =
\frac{1}{8 \pi^2 Q_0}  \left( \frac{f_\pi}{M_{\pi}} \right)^2
\sum_{n_f,s_f}  \frac{\delta(E_f + Q_0 - E_{i})}{4 E_i E_f} R_E.
\end{equation}
In the above equation $R_E$ is written as
\begin{eqnarray}
R_E &=&  \frac{1}{4} {\rm Tr} \left\{ \cO^{\dagger}_{\pi}
\left[ E_f \gamma_0 + P_{fT} \gamma^2 - P_{fz} \gamma^3
+ M_N + (\kappa_p/M_N) \Sigma_z \right] \right. 
\left( A_f + \gamma_5 \asla_f \right) 
\nonumber \\ && \left.\quad \times
 \cO_{\pi}
 \left[ E_i \gamma_0 + P_{iT} \gamma^2 - P_{iz} \gamma^3
 + M_N + (\kappa_p/M_N) \Sigma_z \right]
\left( A_i + \gamma_5 \asla_i \right) \right\} 
\end{eqnarray}
with $P_{i(f)T} = \sqrt{2n_{i(f)} + 1 - s_{i(F)}}$,
and
\begin{eqnarray}
A_{i(f)} &=&1 + \frac{s_i \kappa_p/M_N }
{\sqrt{2n_{i(f)} + 1 - s_{i(F)} + M_N^2} }
\\
\cO_{\pi}
&=&
\gamma_5 \left\{ 
\left[ \cM_1 \frac{1 + \Sigma_z}{2} + \cM_2 \frac{1 - \Sigma_z}{2} \right]
\left[ \gamma_0 Q_0 - \gamma^3 Q_z \right]
\right. \nonumber \\ && \left. \quad
- \left[ \cM_3 \frac{1 + \Sigma_z}{2}
+  \cM_4 \frac{1 - \Sigma_z}{2} \right] \gamma^2 Q_y  \right\} ,
\end{eqnarray}
where
\begin{eqnarray*}
\cM_1 = \cM \left( n_i + \frac{1-s_i}{2}, n_f+\frac{1-s_f}{2} \right),
&&
\cM_2 = \cM \left( n_i - \frac{1+s_i}{2}, n_f - \frac{1+s_f}{2} \right),
\\
\cM_3 = \cM \left( n_i + \frac{1-s_1}{2}, n_f-\frac{1+s_f}{2} \right),
&&
\cM_4 = \cM \left( n_i - \frac{1+s_i}{2}, n_f+ \frac{1-s_f}{2} \right).
\end{eqnarray*}

$R_E$ can be written explicitly as
\begin{equation}
 R_E = \sum_{i<j} M_i M_j R(i,j) .
\end{equation}
with
\begin{eqnarray*}
R(1,1) &=& 
  \frac{1}{2} \left\{ [A_i A_f + s_i s_f (a_i \cdot a_f)]
\left[ (P_{iL} \cdot P_{fL}) (P_{iL}^2 + P_{fL}^2) - P_{iL}^2 P_{fL}^2 
- (M_N + \kappa_p/M_N )^2 Q_L^2 \right] \right. 
\\ &&~
+ s_i s_f   (Q_L \cdot  a_i)(Q_L \cdot  a_f) 
\left[ P_{iL}^2 -  P_{fL}^2 + 2 (M_N + \kappa_p/M_N)^2 \right]
\\ && +  (M_N + \kappa_p/M_N )  A_f s_i 
\left[  ( P_{iL}^2 -  P_{fL}^2)  (Q_0 a_i^z - Q_z a_i^0 )
- 2 Q_L^2 ( E_i a_i^z - P_{iz} a_i^0 ) \right]
\\ && \left.
+ (M_N + \kappa_p/M_N )  A_i s_f \left[
 ( P_{iL}^2 -  P_{fL}^2) (Q_0 a_f^z - Q_z a_f^0 )
- 2 Q_L^2 (E_f  a_f^z - P_{fz} a_f^0  \right] \right\} ,
\\
%
%
R(2,2) &=& 
  \frac{1}{2} \left\{ [A_i A_f + s_i s_f (a_i \cdot a_f)]
\left[ (P_{iL} \cdot P_{fL}) (P_{iL}^2 + P_{fL}^2) - P_{iL}^2 P_{fL}^2 
- ( M_N - \kappa_p/M_N )^2 Q_L^2 \right] \right. 
\\ &&~
+ s_i s_f   (Q_L \cdot  a_i)(Q_L \cdot  a_f) 
\left[ P_{iL}^2 -  P_{fL}^2 + 2 ( M_N - \kappa_p/M_N )^2 \right]
\\ && - ( M_N -  \kappa_p/M_N ) A_f s_i 
\left[  ( P_{iL}^2 -  P_{fL}^2)  (Q_0 a_i^z - Q_z a_i^0 )
- 2 Q_L^2 ( E_i a_i^z - P_{iz} a_i^0 ) \right]
\\ && \left.
 - ( M_N -  \kappa_p/M_N )  A_i s_f \left[
 ( P_{iL}^2 -  P_{fL}^2) (Q_0 a_f^z - Q_z a_f^0 )
- 2 Q_L^2 (E_f a_f^z - P_{fz} a_f^0  \right] \right\} ,
%
\\
R(3,3) &=& \frac{Q_T^2}{2}  \left\{ 
\left[ P_{iL} \cdot P_{fL} + M_N^2 - (\kappa_p/M_N)^2 \right] 
( A_i A_f + s_i s_f a_i \cdot a_f ) 
- s_i s_f ( Q_L \cdot a_i) ( Q_L \cdot a_f )
 \right. \quad\quad\quad
\\ && \left.
- (\kappa_p/M_N) \left[ Q_0 ( A_f s_i a_i^z - A_i s_f a_f^z )
-  Q_z ( A_f s_i 0 a_i^0 - A_i s_f a_f^0 ) \right]
  \right\} .
\\ && - \frac{Q_T^2}{2} M_N  \left\{
 (E_i + E_f) ( A_f s_i a_i^z - A_i s_f a_f^z )
- (P_{iz} + P_{fz}) ( A_f s_i a_i^0 - A_i s_f a_f^0 )
  \right\} ,
\\
R(4,4) &=& \frac{Q_T^2}{2}   \left\{ 
\left[ P_{iL} \cdot P_{fL} + M_N^2 - (\kappa_p/M_N)^2\right] 
( A_i A_f + s_i s_f a_i \cdot a_f ) 
- s_i s_f ( Q_L \cdot a_i) ( Q_L \cdot a_f )
 \right. \quad\quad\quad
\nonumber \\ && \left.
- (\kappa_p/M_N) \left[ Q_0 ( A_f s_i a_i^z - A_i s_f a_f^z )
-  Q_z ( A_f s_i a_i^0 - A_i s_f a_f^0 ) \right]
  \right\} .
\\ && + \frac{Q_T^2}{2} M_N  \left\{
 (E_i + E_f) ( A_f s_i a_i^z - A_i s_f a_f^z )
- (P_{iz} + P_{fz}) ( A_f s_i a_i^0 - A_i s_f a_f^0 )
  \right\} ,
%
%
\\
R(1,2) &=& P_{iT} P_{fT}
\left\{  Q_L^2 \left[ A_i A_f + s_i s_f a_i \cdot a_f \right]
- 2 s_i s_f ( Q_L \cdot a_i ) ( Q_L \cdot a_f )  \right\} ,
%
\\
R(1,3)
 &=& \frac{Q_T}{2} \left\{
- \left[P_{iT} ( P_{fL} \cdot Q_L ) + P_{fT} ( P_{iL} \cdot Q_L ) \right] 
 \left[ A_i A_f + s_i s_f (a_f \cdot a_i) \right] 
\right. \\ && \quad
+ ( P_{iT} + P_{fT} ) s_i s_f (Q_L \cdot a_f)(Q_L \cdot a_i)
 \\ && \left. \quad
+ ( P_{iT} - P_{fT} )  (M_N + \kappa_p/M_N ) 
\left[ A_f s_i ( Q_0 a_i^z - Q_z a_i^0 ) 
- A_i s_f ( Q_0 a_f^s - Q_z a_f^0 )  \right]  \right\} ,
\\
 R(1,4) 
 &=& \frac{Q_T}{2} \left\{
- \left[ P_{iT} (P_{fL} \cdot Q_L) +  P_{fT} ( Q_L \cdot P_{iL}) \right]
\left[ A_i A_f + s_i s_f a_f \cdot a_i \right]  \right.
\\  && \quad
+ ( P_{iT} + P_{fT} ) s_i s_f (Q_L \cdot a_f)(Q_L \cdot a_i)
\\ && \left. \quad \quad 
+ (P_{iT} - P_{fT} ) ( M_N + \kappa_p/M_N ) 
\left[ A_f s_i (Q_0 a_i^z - Q_z a_i^0) 
- A_i s_f (Q_0 a_f^z - Q_z a_f^0) \right] \right\} ,
%
\\ 
R(2,3) &=& R(2,4)
\\ &=&
  \frac{Q_T}{2} \left\{ 
- \left[( P_{iL} \cdot Q_L ) P_{fT} + ( Q_L \cdot P_{fL} ) P_{iT} \right]
\left[ A_i A_f  + s_i s_f  ( a_f \cdot a_i ) \right] 
\right. \\ &&  \quad\quad
+ ( P_{iT} + P_{fT} ) s_i s_f  (Q \cdot a_f)(Q \cdot a_i )
\\ &&   \left. \quad\quad
+ ( P_{iT} - P_{fT} ) (- M_N + \kappa_p/M_N ) 
[ A_f s_i  ( Q_0 a_i^z - Q_z a_i^0 )
 - A_i s_f \left( Q_0 a_f^z  - Q_z a_f^0 \right)]
  \right\} ,
%
%
\\ R(3,4) &=& -  Q_T^2 P_{iT} P_{fT}
\left( A_i A_f + s_i s_f a_i \cdot a_f \right) ,
\end{eqnarray*}
where $P_{i(f)L} \equiv (E_{i(f)},0,0,P_{i(f)z})$ and 
$Q_L  \equiv P_{iL} - P_{fL} = (Q_0, 0, 0, Q_z)$.

\end{document}